          \documentclass[12pt]{article}

\usepackage{amsmath,amssymb}
\def\text{\mbox}
\def\refcite{\cite}
\newcommand\bbasis[2] {b_{#1|#2}}
\newcommand\dbasis[2] {\beta_{#1|#2}}
\newcommand\ddbasis[5]{\beta_{#1#2|#3#4}^{{\scriptscriptstyle(}#5{\scriptscriptstyle)}}}
\newcommand\deltass[2]{\delta_{#1,#2}}

\newtheorem{FFSStheorem}{Theorem}
\newtheorem{FFSScorollary}[FFSStheorem]{Corollary}
\newtheorem{FFSSproposition}[FFSStheorem]{Proposition}
\newtheorem{FFSSremark}[FFSStheorem]{Remark}

\begin{document}

\begin{flushright}
   {\sf NITS-PHY-2012007}\\
   {\sf ZMP-HH/12-21}\\
   {\sf Hamburger$\;$Beitr\"age$\;$zur$\;$Mathematik$\;$Nr.$\;$454}\\[2mm]
   December 2012
\end{flushright}
\vskip 3.5em
\begin{center}
\begin{tabular}c \Large\bf PARTITION FUNCTIONS, MAPPING CLASS GROUPS \\[2mm]
                 \Large\bf AND DRINFELD DOUBLES
\end{tabular}
\end{center}\vskip 2.1em
\begin{center}
   ~Jens Fjelstad\,$^{\,a}$,~
   ~J\"urgen Fuchs\,$^{\,b}$,~
   ~Christoph Schweigert\,$^{\,c}$,~
   ~Carl Stigner\,$^{\,b}$
\end{center}

\vskip 9mm

\begin{center}\it$^a$
   Department of Physics, \ Nanjing University\\
   22 Hankou Road, \ Nanjing,\ 210093\, China
\end{center}
\begin{center}\it$^c$
   Teoretisk fysik, \ Karlstads Universitet\\
   Universitetsgatan 21, \ S\,--\,651\,88\, Karlstad
\end{center}
\begin{center}\it$^c$
   Organisationseinheit Mathematik, \ Universit\"at Hamburg\\
   Bereich Algebra und Zahlentheorie\\
   Bundesstra\ss e 55, \ D\,--\,20\,146\, Hamburg
\end{center}
\vskip 5.3em

\noindent{\sc Abstract}
\\[3pt]
Higher genus partition functions of two-dimensional conformal field theories
have to be invariants under linear actions of mapping class groups. We illustrate
recent results\,\cite{fuSs3,fuSs5} on the construction of such invariants by
concrete expressions obtained for the case of Drinfeld doubles of finite groups.
The results for doubles are independent of
the characteristic of the underlying field, and the general results
do not require any assumptions of semisimplicity.

  \newpage


\section{General results}

To start, let us sketch the incentive for our work from two-dimensional
con\-formal field theory (CFT). The building blocks for CFT correlators
with given chiral data are the conformal blocks,
which form vector bundles with projectively flat
connection over the moduli space of complex curves with marked points.
It is fairly well understood that essential properties of the monodromies of
these connections are encoded in the structure of a ribbon category 
$\mathcal C$. In many applications to physical systems, it is unnatural
to require $\mathcal C$ to be semisimple. Indeed, non-semisimple ribbon 
categories arise in logarithmic conformal field theories,
a topic of much recent activity 
and with many applications, in particular to statistical systems.

In this note we consider linear representations of the mapping class
groups of Riemann surfaces of genus $g$ with $m$ holes, for arbitrary values
of $g$ and $m$, on complex vector spaces that are specific
morphism spaces of a finite ribbon category $\mathcal C$.
Recall that a \emph{finite tensor category} is a rigid monoidal category
with fi\-ni\-te-di\-men\-si\-o\-nal morphism spaces and
finitely many isomorphism classes of simple objects, each of which has a
projective cover, and with every object having finite length;
a \emph{ribbon category} is a rigid monoidal category endowed with
a compatible braiding and twist.

A central idea in quantum field theory is the one of ``summing over all
intermediate states''. Beyond the semisimple setting, one cannot give a
naive mathematical meaning to this by restricting the summation by hand to, 
say, simple or indecomposable or projective objects. The idea is, however, 
appropriately implemented by the categorical notion of a  \emph{coend} 
$\int^{X\in \mathcal C}\!G(X,X)$ of a certain functor 
$G\colon \mathcal C^{\mathrm{op}}\,{\times}\, \mathcal C \,{\to}\, \mathcal D$. 
The coend can be seen as a formalization of the idea to sum over all
possible states, in a way that is consistent with the morphisms in the category.
We have \,\cite{lyub6,KEly}

\begin{FFSStheorem}\label{thm:LC}
Let $\mathcal C$ be a finite ribbon category.
\\
{\em (i)} The coend \\[-25pt]
  \begin{equation}
  L(\mathcal C) := \int^{U\in \mathcal C}\! U^\vee \otimes U
  \label{LC}
  \end{equation}
of the functor $\mathcal C^{\mathrm{op}}\,{\times}\,\mathcal C \,{\to}\, \mathcal C$ 
that acts on objects as $(U,V)\,{\mapsto}\, U^\vee{\otimes}\,V$ exists.
\\[1pt]
{\em (ii)}
The coend $L(\mathcal C)$ carries a natural structure of a Hopf algebra in 
$\mathcal C$. It is endowed with an integral and with a Hopf pairing 
$\varpi_{L(\mathcal C)}:L(\mathcal C) \,{\otimes}\, L(\mathcal C) \,{\to}\, {\mathbf1}$.
\\[1pt]
{\em (iii)}
Via the rigidity of $\mathcal C$, every object of $\mathcal C$ carries a 
natural structure of right comodule over $L(\mathcal C)$.
\end{FFSStheorem}

\begin{FFSSproposition}\label{rhoLX}
{\em \cite[Rem.\,2.3]{fuSs5}}
For $\mathcal C$ a finite ribbon category, every object of $\mathcal C$ carries a 
natural structure of left-right Yetter-Drinfeld module over $L(\mathcal C)$, with 
the right comodule structure being the one of Theorem {\em \ref{thm:LC}(iii)}
and the left module structure obtained via the braiding of $\mathcal C$.
\\
This provides a fully faithful embedding of $\mathcal C$ into
the category of left-right Yetter-Drinfeld modules over $L(\mathcal C)$.
\end{FFSSproposition}

A \emph{factorizable} finite ribbon category is a finite ribbon category 
$\mathcal C$ for which the Hopf pairing $\varpi_{L(\mathcal C)}$ is 
non-degenerate; for such a category the integral of the Hopf algebra 
$L(\mathcal C)$ is two-sided and $L(\mathcal C)$ also has a two-sided cointegral
(see e.g.\ Proposition 5.2.10 and Corollary 5.2.11 of \cite{KEly}).
In the sequel we restrict our attention to a specific subclass of such
categories. We denote by $\Bbbk$ an algebraically closed field of characteristic 
zero and by $H \,{\equiv}\, (H,m,\eta,\Delta,\varepsilon,\mbox{\sc s},R,v)$ a
fi\-ni\-te-di\-men\-si\-o\-nal ribbon Hopf $\Bbbk$-algebra. Here $m$, $\eta$, 
$\Delta$, $\varepsilon$ and $\mbox{\sc s}$ are the product, unit, coproduct, 
counit and antipode of $H$, $R \,{\in}\, H \,{\otimes_{\Bbbk}}\, H$ is the 
R-matrix and $v \,{\in}\, H$ the ribbon element of $H$. Such a Hopf algebra
has, uniquely up to scalars, a non-zero left integral $\Lambda\,{\in}\, H$ 
and a right cointegral $\lambda\,{\in}\, H^*_{}$; the antipode of $H$ is
invertible. We denote by $Q \,{:=}\, R_{21}\,{\cdot}\, R \,{\in}\, H 
\,{\otimes_{\Bbbk}}\, H$ the monodromy matrix and by $f_Q \,{=}\, (d_H 
\,{\otimes}\, \mbox{\sl id}_H) \,{\circ}\, (\mbox{\sl id}_{H^*_{}} \,{\otimes}\, Q)
\colon H^*_{} \,{\to}\, H$ the \emph{Drinfeld map}.

We also assume that the Hopf algebra $H$ is \emph{factorizable}, meaning that the
Drinfeld map $f_Q$ is invertible. In this case the integral $\Lambda$ is two-sided
(in other words, the Hopf algebra $H$ is \emph{unimodular}),
and the Drinfeld map sends the cointegral $\lambda$ to a non-zero multiple of 
the integral. We can (and do) normalize them such that $\lambda\,{\circ}\,\Lambda 
\,{=}\, 1$ and $f_Q(\lambda) \,{=}\, \Lambda$.

Complex Hopf algebras with structure very close to the one considered
here arise in the description of representation categories of
chiral algebras that are not semisimple, see e.g.\
\cite{fgst,fgst4,naTs2,tswo,rugw}.

\medskip

Now denote by $H\text{-Mod}$ the category of (fi\-ni\-te-di\-men\-si\-o\-nal, 
left) $H$-mo\-du\-les. The following is
an immediate consequence of well-known results:

\begin{FFSSproposition}
The category $H\text{-Mod}$ carries a natural structure of a factorizable 
finite ribbon category.
Specifically, the tensor product of $H$-Mod is obtained by pull-back of the
coproduct $\Delta$ of $H$, the left and right dualities come from the antipode
of $H$ and its inverse, the braiding
comes from the R-matrix $R$, and the twist comes from the ribbon element $v$.
\end{FFSSproposition}

The category $H\mbox{-}\mathrm{Bimod}$ of fi\-ni\-te-di\-men\-si\-o\-nal
$H$-bimodules can be treated very much in the same vein, giving (see 
\refcite{fuSs3})

\begin{FFSScorollary}
The category $H\mbox{-}\mathrm{Bimod}$ of fi\-ni\-te-di\-men\-si\-o\-nal
$H$-bimodules has a natural structure of a factorizable finite ribbon category.
\end{FFSScorollary}

\begin{FFSSremark}
(i)
It is worth pointing out that the relevant monoidal structure is \emph{not} the 
one for which the tensor product is taken over $H$ (and thus treats the Hopf
algebra $H$ just as an algebra).
Braidings for that other monoidal structure are discussed in \cite{agcm}.
\\
(ii)
We can use the ribbon structure on $H$ to
equip the category $H\mbox{-}\mathrm{Bimod}$ (endowed with the relevant
monoidal structure) with a natural structure of a ribbon category. 
To this end, we can use the simultaneous left 
action and right action of either the R-matrix and the ribbon element, or else 
the inverse R-matrix and inverse ribbon element.  Altogether this results in 
four structures of ribbon category on 
$H\mbox{-}\mathrm{Bimod}$; these are pairwise isomorphic. For our purpose it is 
crucial to take one of the two ribbon structures in which mutually inverse 
elements are used on the left and on the right; for concrete expressions 
see the formulas (3.3) and (4.19) of \refcite{fuSs3}. This is precisely 
the ribbon structure that makes $H\mbox{-}\mathrm{Bimod}$, for factorizable $H$, 
braided equivalent to the Drinfeld center of $H\text{-Mod}$.
\end{FFSSremark}

As shown in Appendix A.2 of \refcite{fuSs3}, there is an equivalence
  \begin{equation}
  {H\text{-Mod}}^{\rm rev} \;{\boxtimes}\;H\text{-Mod}
  \stackrel\simeq\longrightarrow H\mbox{-}\mathrm{Bimod}
  \label{HHequivHb}
  \end{equation}
of ribbon categories, where $\boxtimes$ is the Deligne tensor product of abelian 
categories and the \emph{reverse} category $\mathcal C^{\rm rev}$ of a ribbon
category $\mathcal C$ is obtained from $\mathcal C$ by inverting
the braiding and the twist isomorphisms.
We use this equivalence to tacitly identify these two categories. In particular,
we think of the coend $L({H\text{-Mod}}^{\rm rev}{\boxtimes}H\text{-Mod})$, as 
introduced in \eqref{LC}, as an $H$-bimodule. We denote this bimodule by $K$; 
its structure as an $H$-bimodule and as a Hopf algebra in $H\mbox{-}\mathrm{Bimod}$ 
are described in detail in Appendix A.3 of \refcite{fuSs3}.

The simplest partition function in the semisimple case is the charge conjugation
modular invariant; this is of Peter-Weyl form and thus a direct sum of terms
in bijection with the isomorphism classes of simple $H$-modules.
In the non-semisimple case, the direct sum gets again generalized to a coend,
this time of a functor $G^H\colon H\text{-Mod}^{\mathrm{op}}
\,{\times}\,H\text{-Mod} \,{\to}\, H\mbox{-}\mathrm{Bimod}$. The functor $G^H$ in 
question is the one obtained by composing the func\-tor $H\text{-Mod}^{\mathrm{op}}
\,{\times}\,H\text{-Mod} \,{\to}\,{H\text{-Mod}}^{\rm rev}\,{\boxtimes}\,H\text{-Mod}$
that acts on objects as $U\,{\times}\,V \,{\mapsto}\, U^\vee\,{\boxtimes}\,V$
with the equivalence \eqref{HHequivHb}.
The coend of $G^H$ implements in a consistent manner the idea to pair left and
right movers in the form of charge conjugation on \emph{all} representations.

\begin{FFSSproposition}
{\em \cite[Propos.\,A.3]{fuSs3}}
The coend $F$ of the functor $G^H\colon$ $H\text{-Mod}^{\mathrm{op}}\,{\times}
\,H\text{-Mod} \,{\to}\, H\mbox{-}\mathrm{Bimod}$ exists. Its
underlying object -- also denoted by $F$ -- is the \emph{coregular bimodule},
i.e.\ the vector space $H^*_{}$ dual to $H$ endowed with the duals of the left 
and right regular actions of $H$ on itself.
\end{FFSSproposition}

Similarly as the coend $K$, also the coend $F$ turns out to carry important 
additional algebraic structure internal to $H\mbox{-}\mathrm{Bimod}$:

\begin{FFSStheorem}
{\em \cite[Theorem\,2]{fuSs4}}
The coend $F$ carries a natural structure of a commutative symmetric Frobenius 
algebra with trivial twist in the ribbon category $H\mbox{-}\mathrm{Bimod}$.
The algebra and coalgebra structural maps are given by
  \begin{equation}
  \begin{array}{ll}
  m_F := {\Delta^{\phantom:}}^{\!\!\!*_{}} \,, \qquad
  \eta_F := \varepsilon^* \,, \qquad
  \varepsilon_F := {\Lambda^{\phantom:}}^{\!\!\!*} \qquad{\rm and}\qquad
  \\[2mm]
  \Delta_F := {[ (\mbox{\sl id}_H \,{\otimes}\, (\lambda\,{\circ}\, m)) \,{\circ}\,
  (\mbox{\sl id}_H\,{\otimes}\,\mbox{\sc s}\,{\otimes}\,\mbox{\sl id}_H)
  \,{\circ}\, (\Delta\,{\otimes}\,\mbox{\sl id}_H) ]}^* ,
  \end{array}
  \label{Ffro}
  \end{equation}
with $\Lambda$ and $\lambda$ the integral and cointegral of $H$, respectively.
\end{FFSStheorem}

We now turn to the action of mapping class groups. The results in
\refcite{lyub6} imply that for any triple of non-negative integers
$g$, $p$ and $q$, the morphism space $V_{g;p,q} \,{=}\, \mathrm{Hom}_{H|H}
(K^{\otimes g}\,{\otimes}\, F^{\otimes q}, F^{\otimes p})$ naturally carries 
a projective representation $\pi_{g;p,q}$ of the subgroup Map$_{g;p,q}$ of 
the mapping class group of Riemann surfaces of genus $g$ with $p\,{+}\,q$ 
holes that leaves two selected subsets of sizes $p$ and $q$ separately invariant.
For details see Propos.\ 2.4 and Remark 2.6 of \refcite{fuSs5}.
$\pi_{g;p,q}$ is in fact a genuine linear representation, see Remark 5.5
of \refcite{fuSs3}.

In applications to CFT, $F$ is a candidate for the bulk state space,
and the spaces $V_{g;p,q}$ play the role of genus-$g$ conformal blocks with $p$
incoming and $q$ outgoing insertions of the bulk state space. The corresponding
correlation functions are specific elements in those spaces that have to be
invariant under the mapping class group action and be compatible with sewing.

We now present elements $\mathrm{Cor}_{g;p,q} \,{\in}\, V_{g;p,q}$ that are 
candidates for these correlation functions. Denote by $m_F^{(r)}$ and 
$\Delta_F^{(r)}$ multiple products and coproducts of $F$, respectively, and 
by $\rho^K_F$ the natural action (see Propos.\ \ref{rhoLX}) of the Hopf 
algebra $K \,{\in}\, H\mbox{-}\mathrm{Bimod}$ on the $H$-bimodule $F$, and set
  \begin{equation}
  \begin{array}{l}
  \mathrm{Cor}_{0;1,1} := \mbox{\sl id}_F \,, \qquad
  \mathrm{Cor}_{1;1,1} := m_F \circ (\rho^K_F \,{\otimes}\, \mbox{\sl id}_F)
    \circ (\mbox{\sl id}_K \,{\otimes}\, \Delta_F) \,,
  \\[4pt]
  \mathrm{Cor}_{g;1,1} := \mathrm{Cor}_{1;1,1} \circ (\mbox{\sl id}_K \,{\otimes}\,
    \mathrm{Cor}_{g-1;1,1}) \quad~ {\rm for}~~ g\,{>}\,1 \,,
  \\[4pt]
  \mathrm{Cor}_{g;p,q} := \Delta_F^{(p)} \circ \mathrm{Cor}_{g;1,1} \circ
  \big( \mbox{\sl id}^{}_{K^{\otimes g}_{}} \,{\otimes}\, m_F^{(q)} \big) \,.
  \end{array}
  \label{corrs}
  \end{equation}

\begin{FFSStheorem} \label{fuSs5main}
The element $\mathrm{Cor}_{g;p,q}$ of $V_{g;p,q}$
is invariant under the action $\pi_{g;p,q}$ of the group Map$_{g;p,q}$.
\end{FFSStheorem}

This statement is the main result of \refcite{fuSs5}.
It provides yet another instance of a surprising conspiracy of algebraic
structures -- in the case at hand,
finite-dimensional factorizable Hopf algebras -- and geometric
ones -- here, the fundamental groups of moduli spaces of complex curves.
We find it remarkable that semisimplicity of the categories involved
is needed neither for producing mapping class group representations nor
for the construction of physically motivated invariants.


\section{Drinfeld doubles}

Associated to any finite group $G$ there is the quasitriangular Hopf algebra 
$\mathcal DG$, called the (Drinfeld) double of $G$. As a vector space, 
$\mathcal DG \,{=}\, \Bbbk(G) \,{\otimes_{\Bbbk}}\, \Bbbk[G]$ with 
$\Bbbk(G)$ the algebra of functions on $G$ and $\Bbbk[G]$ the group algebra
over an (algebraically closed) field $\Bbbk$. From here on we do not assume 
any longer that $\Bbbk$ has characteristic zero, and thus do not require 
$\mathcal DG$ to be semisimple. A natural basis of $\mathcal DG$ is given 
by $\bbasis gx \,{:=}\, \delta_g \,{\otimes}\, b_x$ with $g,x\,{\in}\, G$, 
where $\delta_g$ and $b_x$ are the usual natural bases of $\Bbbk(G)$ and 
$\Bbbk[G]$. In terms of this basis the Hopf algebra structure of
$\mathcal DG$ reads
  \begin{eqnarray} &&
  \bbasis gx\cdot \bbasis hy = \deltass g{xhx^{-1}}\, \bbasis g{xy} \,,
  \quad
  \eta(1) = \bbasis 1e = \mbox{$\displaystyle\sum$}_{g\in G}\, \bbasis g e \,,
  \quad
  \varepsilon(\bbasis gx) = \deltass ge \,,
  \nonumber \\[3pt] &&
  \Delta(\bbasis gx) = \mbox{$\displaystyle\sum$}_{h\in G}\, \bbasis hx \,{\otimes}\, \bbasis {h^{-1} g}x \,,
  \quad~
  \mbox{\sc s}(\bbasis gx) = \bbasis {x^{-1} g^{-1} x}{x^{-1}} \,.
  \end{eqnarray}
Note that $\mbox{\sc s}^2 \,{=}\, \mbox{\sl id}_{\mathcal DG}$.
The Hopf algebra $\mathcal DG$ is factorizable quasitriangular, with R-matrix and
associated monodromy matrix given by
  \begin{equation}
  R = \mbox{$\displaystyle\sum$}_{g\in G}\, \bbasis ge \,{\otimes}\, \bbasis 1g
  \qquad \mathrm{and}\qquad
  Q = \mbox{$\displaystyle\sum$}_{g,h\in G}\, \bbasis hg \,{\otimes}\, \bbasis g{g^{-1} hg} \,.
  \end{equation}
$\mathcal DG$ is also ribbon, with ribbon element $\nu \,{=}\, \sum_{g\in G}\bbasis g{g^{-1}}$,
and it has a two-sided integral and a two-sided cointegral, given (upon choice
of normalizations) by $\Lambda \,{=}\, \sum_{g}\bbasis e g$ and by $\lambda \,{=}\, 
\sum_{g} \dbasis ge \,{\in}\,\mathcal DG^*_{}$, where $\{\dbasis gx\}$ is the
basis of the dual space $\mathcal DG^*_{}$ dual to $\{\bbasis gx\}$.
The integral satisfies $\varepsilon \,{\circ}\, \Lambda \,{=}\, |G|$; accordingly,
by Maschke's theorem the category $\mathcal DG\text{-Mod}$ of left $\mathcal DG$-modules
is semisimple iff the characteristic of $\Bbbk$ does not divide the order $|G|$.


\section{Invariants from Drinfeld doubles}

Let us apply some of the general results from Section 1 to the factorizable 
Hopf algebra $H \,{=}\, \mathcal DG$. First, the Hopf algebra $K$ in
$\mathcal DG\mbox{-}\mathrm{Bimod}$ is $(\mathcal DG^*_{}) \,{\otimes_{\Bbbk}}\,
(\mathcal DG^*_{})$ as a vector space, with the coadjoint left and right 
$\mathcal DG$-actions on the first and second tensor factor, respectively,
and its Hopf algebra structure is easily expressed in 
terms of the dual basis $\{\dbasis gx\}$.
The Hopf algebra structure of $K \,{\in}\,\mathcal DG\mbox{-}\mathrm{Bimod}$,
expressed in terms of the dual basis $\{\dbasis gx\}$, is given by
  \begin{eqnarray} &&
  (\dbasis gx{\otimes}\dbasis hy)\cdot(\dbasis pu{\otimes}\dbasis qv)
  = \deltass {px}{up}\, \deltass {qvy}{vqv}\, \dbasis {pg}u \,{\otimes}\,
    \dbasis {v^{-1} qvhv^{-1} q^{-1} vq}v \,,
  \nonumber \\[4pt]&&
  \Delta_K(\dbasis gx\,{\otimes}\,\dbasis hy) = \mbox{$\displaystyle\sum$}_{u,v\in G}\,
  \dbasis gv\,{\otimes}\, \dbasis {u^{-1} hu}{u^{-1} y} \,{\otimes}\,
    \dbasis {v^{-1} gv}{v^{-1} x} \,{\otimes}\, \dbasis hu \,,
  \nonumber \\[2pt]&&
  1_K = \mbox{$\displaystyle\sum$}_{x,y\in G}\, \dbasis ex \,{\otimes}\, \dbasis ey \,, \quad
  \varepsilon_K(\dbasis gx\,{\otimes}\,\dbasis hy) = \deltass xe\,\deltass ye \,, \quad
  \nonumber \\[3pt]&&
  \mbox{\sc s}_K(\dbasis gx{\otimes}\dbasis hy)
    = \dbasis {x^{-1} g^{-1} x}{x^{-1} g^{-1} x^{-1} gx}
      \otimes \dbasis {h^{-1} y^{-1} h^{-1} yh}{h^{-1} y^{-1} h} \,.
  \end{eqnarray}
The Frobenius algebra $F$ in $\mathcal DG\mbox{-}\mathrm{Bimod}$ is 
$\mathcal DG^*_{}$ as a vector space, with the coregular left and right 
$\mathcal DG$-actions.  The Frobenius algebra structure is obtained by 
specializing formula \eqref{Ffro} to the present situation; we find
  \begin{equation}
  \begin{array}{lll}
  \eta_F^{} = \mbox{$\displaystyle\sum$}_{x\in G}\, \dbasis ex \,, \quad&~&
  m_F^{}(\dbasis gx,\dbasis hy) = \deltass xy\, \dbasis{hg}x \,,
  \\[3pt]
  \varepsilon_F^{}(\dbasis gx) = \deltass ge \,,&&
  \Delta_F^{}(\dbasis gx) = \mbox{$\displaystyle\sum$}_{h\in G}\,
  \dbasis{h^{-1} g}x \,{\otimes}\, \dbasis hx \,.
  \end{array}
  \label{DGF}
  \end{equation}

Now consider the action $\rho^K_F$ of the Hopf algebra $K$ on the
Frobenius algebra $F$, which is a linear map from
$(\mathcal DG^*_{})_{}^{\otimes_{\Bbbk} 3}$ to $\mathcal DG^*_{}$. We obtain
  \begin{equation}
  \rho^K_F(\dbasis gx{\otimes}\dbasis hy{\otimes}\dbasis kz)
   = \deltass z{xzy}\, \deltass kx\, \dbasis {g^{-1} xg}{g^{-1} zy^{-1} hy} \,.
  \label{DGrho}
  \end{equation}
We can now insert the specific expressions \eqref{DGF} and \eqref{DGrho}
into the general formula \eqref{corrs}. We restrict our attention to the case 
$p \,{=}\, 1 \,{=}\, q$; the extension to general
values of $p$ and $q$ is straightforward. At genus 1 we have
  \begin{equation}
  \mathrm{Cor}_{1;1,1} (\dbasis gx{\otimes}\dbasis hy{\otimes}\dbasis kz)
  = \deltass {y^{-1}}{z^{-1} xz}\, \deltass {y^{-1} h y}{z^{-1} gz}\, \dbasis {k [x,g]}z \,,
  \end{equation}
with $[x,g] \,{=}\, x^{-1} g^{-1} x g$ the group commutator.
By iteration we arrive at
  \begin{equation}
  \mathrm{Cor}_{n;1,1} (\ddbasis ghxy n {\otimes} \dbasis kz)
  = \mbox{$\displaystyle\prod$}_{i=1}^n\, \deltass{y^{-1}_i}{z^{-1} x_i^{}z}\,
  \deltass{y^{-1}_i h_i^{}y_i^{}}{z^{-1} g_i^{} z}\,
  \dbasis {k[x_n^{},g_n^{}]\cdots [x_1^{},g_1^{}]}{z} \,,~
  \end{equation}
where we introduced the short-hand notation
  \begin{equation}
  \ddbasis ghxy n := \dbasis {g_1^{}}{x_1^{}}\otimes\dbasis {h_1^{}}{y_1^{}}\otimes
  \dbasis{g_2^{}}{x_2^{}}\otimes\dbasis{h_2^{}}{y_2^{}}
  \otimes\,\cdots\,\otimes\dbasis{g_n^{}}{x_n^{}}\otimes\dbasis{h_n^{}}{y_n^{}} \,.
  \label{Corrs4dble}
  \end{equation}

According to Theorem \ref{fuSs5main} the morphisms \eqref{Corrs4dble} are
invariant under the action of Map$_{g;p,q}$ if $\Bbbk$ has characteristic zero.
But we actually expect that this remains true for general characteristic,
including the case that the cha\-racteristic divides the order
of $G$ so that $\mathcal DG$ is non-semisimple.

Indeed, we have verified that the torus partition function $\mathrm{Cor}_{1;0,0}$ 
is modular invariant irrespective of the characteristic of $\Bbbk$.
The action of the S- and T-transformation on $\mathrm{Cor}_{1;0,0}$ is given by
precomposition with the morphisms $S_K$ and $T_K$ depicted in (4.3) of \cite{fuSs5},
   which in the case at hand read
  \begin{equation}
  T_K(\dbasis gx{\otimes}\dbasis hy) = \dbasis g{gx}\otimes\dbasis h{h^{-1} y}
  \end{equation}
and
  \begin{equation}
  S_K(\dbasis gx{\otimes}\dbasis hy)
  = \dbasis {g^{-1} x g}{g^{-1}}\otimes\dbasis {y^{-1}}{y^{-1} hy} \,.
  \end{equation}
Checking mapping class group invariance of correlators with field insertions
and at higher genus amounts to specializing various expressions from
\cite[Sect.\,5]{fuSs5}; this would be straightforward, but some of the 
computations involved tend to be lengthy.

\vskip 2em {\small

 \noindent{\sc Acknowledgments:}
This work is supported in part by the ESF Research Networking Programme 
``Interactions of Low-Dimensional Topology and Geometry with Mathematical 
Physics''.  
The collaboration between JFj, JFu and CSt is also supported by the VR
Swedish Research Links Programme under project no.\ 348-2008-6049;
JFu is in addition supported by VR under project no.\ 621-2009-3993.
JFj is supported by 985 Project Grants from the Chinese Ministry of Education, 
by the Priority Academic Program Development of Jiangsu Higher Education 
Institutions (PAPD), and by the China Science Postdoc grant 2011M500887.
CSc is partially supported by the Collaborative Research Centre 676 ``Particles,
Strings and the Early Universe - the Structure of Matter and Space-Time'' and
by the DFG Priority Programme 1388 ``Representation Theory''.
}



\bigskip

\begin{thebibliography}{12}
\bibitem{agcm} A.L.\ Agore, S.\ Caenepeel, and G.\ Militaru, e-print {math.QA/1108.2575}.
\bibitem{fgst}  B.L.\ Feigin, A.M.\ Gainutdinov, A.M.\ Semikhatov, and I.Yu.\ Tipunin,
           Com\-mun.\ Math.\ Phys.\ 265 (2006) 47.
\bibitem{fgst4} B.L.\ Feigin, A.M.\ Gainutdinov, A.M.\ Semikhatov, and I.Yu.\ Tipunin,
           J.\ Math.\ Phys.\ 48 (2007) 032303.
\bibitem{fuSs3} J.\ Fuchs, C.\ Schweigert, and C.\ Stigner, J.\ of Algebra 363 (2012) 29.
\bibitem{fuSs4} J.\ Fuchs, C.\ Schweigert, and C.\ Stigner, in: {\it Strings, Gauge Fields, 
           and the Geometry Behind.\ The Legacy of Maximilian Kreuzer} (A.\ Rebhan et al, eds.;
           World Scientific, Singapore 2012), p.\ 289.
\bibitem{fuSs5} J.\ Fuchs, C.\ Schweigert, and C.\ Stigner, e-print math.QA/1207.6863.
\bibitem{kerl5} T.\ Kerler, in: {\it Quantum Invariants and Low-Dimensional Topology}
           (J.E.\ Andersen et al., eds; Dekker, New York 1997), p.\ 503.
\bibitem{KEly} T.\ Kerler and V.V.\ Lyubashenko,
           {\it Non-Semisimple Topological Quantum Field Theories for 3-Manifolds with Corners}
           (Springer Verlag, New York 2001).
\bibitem{lyub6} V.V.\ Lyubashenko, Com\-mun.\ Math.\ Phys.\ 172 (1995) 467.
\bibitem{lyub8} V.V.\ Lyubashenko, J.\ of Pure and Applied Algebra 98 (1995) 279.
\bibitem{naTs2} K.\ Nagatomo and A.\ Tsuchiya, in: Exploring New Structures and Natural
            Constructions in Mathematical Physics (K.\ Hasegawa et al., eds.;
            Math.\ Soc.\ of Japan, Tokyo 2010), p.\ 1.
\bibitem{rugw} I.\ Runkel, M.R.\ Gaberdiel, and S.\ Wood, e-print {1201.6273}.
\bibitem{tswo}  A.\ Tsuchiya and S.\ Wood, e-print {1201.0419}.
\end{thebibliography}
\end{document}